\documentclass[%
 reprint,
 amsmath,amssymb,
 aps,
 author-year,
]{revtex4-2}

\usepackage{graphicx}
\usepackage{dcolumn}
\usepackage{bm}
\usepackage{amsmath}
\usepackage{mathtools}
\usepackage{ulem}
\usepackage{xcolor}

\usepackage{float}

\begin{document}

\preprint{APS/123-QED}

\title{Electrothermal Dynamics of Cold Front in Impure Tokamak Plasmas}

\author{S. Oshiro}
\author{A. Matsuyama}%
\author{Y. Nakamura}

\affiliation{Graduate School of Energy Science, Kyoto University, Gokasho, Uji, Kyoto 611-0011, Japan \\ 
\normalfont Author to whom correspondence should be addressed: oshiro.shun.66j@st.kyoto-u.ac.jp}


\begin{abstract}
Current density perturbations induced by radiative collapse, which is a possible mechanism governing tokamak plasma disruptions, have been investigated using a reaction-diffusion model. The reaction term of the current diffusion equation, which depends on the first and second radial derivatives of the electrical resistivity profile, produces a strong disturbance in the current density profile in a narrow layer of the cold front. While the current density locally increases in the region where the electron temperature gradient is steep, it decreases behind the cold front in the region where the electron temperature profile exhibits a pronounced concave-down curvature. The electrothermal dynamics driven by such a shape of the current density perturbation and the competition between Ohmic heating and impurity radiation are simulated by the tokamak transport code INDEX. 
\end{abstract}

\maketitle

%
%
%
%
%
\section{Introduction}
Disruption is a major threat to the operation of reactor-grade tokamak devices. A release of thermal and electromagnetic energies during disruption causes thermal loads on the plasma-facing components and electromagnetic loads on in-vessel components and the vacuum vessel \cite{Lehnen(2015)}. Natural disruptions can occur due to various trigger mechanisms \cite{Hender(2007)}, but physically interesting and practically important one is impurity radiation. An importance of impurities arises from the fact that the power loss due to
line radiation is essential for the transition to a resistive plasma, which results in rapid current decay. Impurities also play an important role in mitigating the disruption loads. By injecting massive material into core plasmas and facilitating the conversion of stored energy into low-energy line radiation, it is possible to avoid the localization of heat load and excessive electromagnetic force \cite{Lehnen(2015)}. The feasibility of such schemes has been tested by means of Massive Gas Injection (MGI) and Shattered Pellet Injection (SPI) in the present experiments \cite{Bakhtiari_2011,Commaux_2016, Heinrich_2025}. \par
Massive impurities cool the plasma edge, causing disturbances in the current density profile and leading to global magnetohydrodynamic (MHD) instabilities during disruptions. Earlier numerical studies that addressed tearing instabilities driven by impurity radiation \cite{Drake_sim,Morozov_2005,NR1,white_2015,Teng_2018} show that the perturbations of the current density are induced by local radiative collapse. Perturbations in the electron temperature ($T_{\mathrm{e}}$) and parallel current density ($j$) are coupled with each other, often yielding electrothermal instability in resistive plasmas \cite{Furth(1970),Ryutov}. When a cold front is formed by local radiative collapse, significant disturbance in the $j$ profile occurs, which is localized within a narrow layer of the cold front and can nonlinearly destabilize ideal and/or resistive MHD modes \cite{Drake_sim, Morozov_2005,NR1}. This mechanism is thought to be a possible explanation for the trigger of global reconnection events after massive material injection by the MGI and SPI \cite{Fable_2016,Linder_2020,Bodner_2025}. Such an importance of the current density perturbations motivates us to investigate the electrothermal dynamics of the cold-front formation. \par
The aim of this work is to gain insights into the cold front dynamics from a one-dimensional tokamak transport model. Similar models have been used in the literatures on natural disruptions \cite{Turner_1982} and disruption mitigation \cite{INDEX, Fable_2016, Vallhagen_2025}. We consider the time evolution of the radial profiles of $T_\mathrm{e}$ and $j$ when an impurity source is applied with a radially uniform profile. Enhanced thermal conduction is also introduced to investigate its effects on thermal stability. Although these simulation conditions are not intended to represent any specific experiment, they are useful for identifying the key parameters governing cold front formation. As discussed in the context of Radiative Condensation Instability (RCI) \cite{Aranson}, a cold front is formed when radiation dominates compared to heat transport from upstream and is caused by two adjacent points in space bifurcating into different stable solutions of thermal power balance. The formation of a cold front can also be regarded as a manifestation of bistability in reaction-diffusion systems \cite{RDE}. This implies that the cold front can  propagate as a traveling wave. The behavior of the reaction-diffusion system is reminiscent of the cold front propagation considered in fusion plasmas.\par 
The idea of this paper is to investigate the thermal collapse given by a one-dimensional transport equation on the basis of analytical solutions of a reaction-diffusion system. Both the thermal transport and current diffusion equations are treated as one-dimensional reaction-diffusion equations. It is shown that strongly localized perturbations to the $j$ profile do not stem from the radial diffusion term but from the reaction term, which depends on the first and second radial derivatives of the electrical resistivity profile. The current density increases locally in the region where the electron temperature gradient is steep, whereas it decreases locally in the region where the electron temperature profile exhibits a pronounced concave-down curvature. In the respective regions, the current density perturbation can induce local changes in the electron temperature, including Ohmic reheating and radiative cooling. We illustrate such numerical simulations of radiative collapse using the tokamak transport code INDEX \cite{INDEX}.\par
The remainder of this paper is organized as follows. In Sec. \ref{sec2}, we introduce a reaction-diffusion equation to describe the radiative collapse in impure tokamak plasmas, where impurity radiation dominates over Ohmic heating. In Sec. \ref{sec3}, the radial current diffusion is analyzed for a sigmoid-shaped electron temperature profile that models a cold front, which clarifies the electrothermal response to local radiative collapse driven by strong impurity radiation. In Sec. \ref{INDEX_results}, the numerical simulations of the INDEX code are presented. Finally, Sec. \ref{dis_con} presents the conclusions of this study with several remarks on the connection between the current analysis and experimental observations. 
%
%
%
%
%
%
\section{Nonlinear Reaction-Diffusion Model of Thermal Transport Equation}
\label{sec2}
We consider a reduced model of the radiative collapse in a cylindrical tokamak geometry. A one-dimensional thermal transport equation with respect to the radial coordinate in a cylindrical geometry is considered, where we focus on the contributions of thermal conduction, Ohmic heating $\eta j^{2}$, and impurity radiation $nn_{\mathrm{z}}L$ to the time evolution of the electron temperature $T_{\mathrm{e}}$. Here, $\eta$ is Spitzer resistivity, $j$ is the current density parallel to the magnetic field, $n_{\mathrm{z}}$ is the impurity density and $L$ is the radiative cooling rate. In this section, the radiative cooling rate $L$ is assumed to follow coronal equilibrium; thus, $L=L(T_{\mathrm{e}})$. Although the electron density $n$ should change depending on the impurity charge state, it is assumed to be constant to simplify the following derivation. We begin with the following thermal transport equation:
\begin{equation}
  3\dfrac{\partial T}{\partial t}=\dfrac{1}{r}\dfrac{\partial}{\partial r}\left( r\chi \dfrac{\partial T}{\partial r} \right)+\dfrac{1}{n}\left( \eta j^{2}-nn_{\mathrm{z}}L \right),
  \label{TTE}
\end{equation}
where we assume that the electron temperature is equal to the ion temperature, $T=T_{\mathrm{e}}=T_{\mathrm{i}}$. Assuming that the thermal diffusivity $\chi$ is uniform and independent of the temperature $T$, we obtain the following nonlinear reaction-diffusion (NRD) model from Eq. (\ref{TTE}):
\begin{equation}
  \dfrac{\partial T}{\partial t}=\chi \dfrac{\partial^{2} T}{\partial r^{2}}+\dfrac{\chi}{r}\dfrac{\partial T}{\partial r}+ f\left( T \right),
  \label{NRD_model}
\end{equation}
where $f\left( T \right)=\eta j^{2}/n-n_{\mathrm{z}}L$ is the reaction term. In Eq. (\ref{NRD_model}), the factor of 3 is dropped to simplify the notation.\par
The formation of the cold front can arise from either the gradient of the impurity density or the temperature derivative of the radiative cooling rate. When thermal diffusion is weak, Eq. (\ref{NRD_model}) becomes $d T/d t = f\left( T \right)$, which is in the form like a logistic equation describing the population growth \cite{RDE}. We also assume that the impurity radiation dominates over Ohmic heating. These assumptions are removed once a steep cold front, which induces a strong current density perturbation, has formed, as discussed in the following section. By taking the derivative with respect to $r$ under these assumptions, one obtains the time evolution of the electron temperature gradient as
\begin{equation}
 \dfrac{\partial T_{r}}{\partial t}=-\dfrac{\partial n_{z}}{\partial r}L-n_{z}T_{r}\dfrac{\partial L}{\partial T},
 \label{gradient_eq}
\end{equation}
where $T_{r} \coloneqq \dfrac{\partial T}{\partial r}$. We consider an initially flat, monotonic temperature gradient profile with $T_{r} < 0$ and a flat impurity density profile, as a premise for the following discussion. For neon, the radiative cooling rate curve has a pronounced peak around 30 eV under coronal equilibrium. The temperature gradient develops below 120 eV, where $\partial L/\partial T<0$, as the temperature decreases down to 30 eV. \par 
\begin{figure}
  \includegraphics[trim=0cm 0cm 0cm 0cm, clip, width=0.4\textwidth]{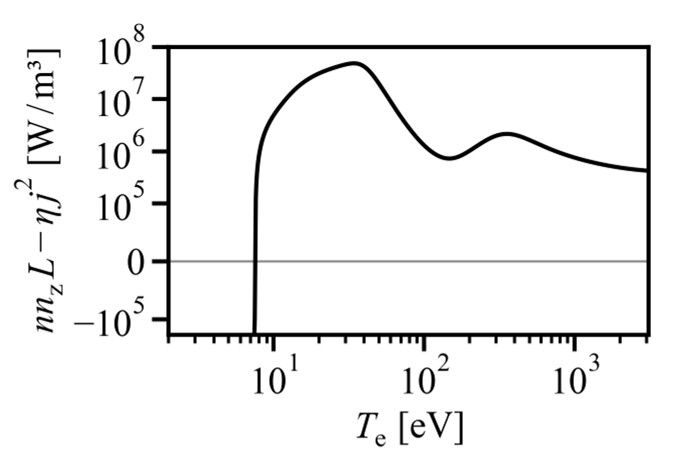}
  \caption{Relationship between the net radiation loss $nn_{\mathrm{z}}L - \eta j^{2}$ and electron temperature, with a fixed current density of $j = 0.3\ \mathrm{MA/m^{3}}$, neon density of $n_{\mathrm{z}} = 10^{19}\ \mathrm{m^{-3}}$, and deuterium density $n_{\mathrm{D}} = 5 \times 10^{19}\ \mathrm{m^{-3}}$.}
  \label{1e19_L-H_T}
\end{figure}
Figure \ref{1e19_L-H_T} shows the electron temperature dependence of the reaction term in Eq. (\ref{NRD_model}) of the coronal equilibrium for impurity charge states. The rate coefficients are taken from the OpenADAS database \cite{ADAS}. Here, the deuterium density is $5 \times 10^{19}\ \mathrm{m}^{-3}$, $j=0.3\ \mathrm{MA/m^{2}}$, and the neon atom density is $n_{\mathrm{z}} = 10^{19}\ \mathrm{m}^{-3}$. The net radiation loss exhibits a peak at around 30 eV. The root of the power balance $\eta j^{2}= nn_{\mathrm{z}}L$ at approximately 7.5 eV is stable. For Ohmic plasmas with a large amount of impurity, only one stable root tends to appear at low temperature because of the electron temperature dependence of the Spitzer resistivity, $\eta \propto T_{\mathrm{e}}^{-3/2}$. Nonetheless, a cold front can be obeserved because there is a difference of nearly two orders of magnitude in radiation power between the peak value at 30 eV and the minimum at 120 eV. On the timescale of fast radiative collapse -- even that exact power balance is not maintained at that temperature -- the $T_{h} \simeq$ 120 eV plays a key role as a \textit{quasi-stable} root. Therefore, on the fast timescale, thermal instability occurs. In previous studies of thermal instability \cite{RCI1,RCI2,Ohayabu,Stacey,Neuhauser,Drake}, thermal conduction competes with the growth of impurity radiation in the temperature range where $\partial L/\partial T_{\mathrm{e}}<0$. From the analysis of Eq. (\ref{gradient_eq}) and linear theory \cite{RCI1,RCI2,Ohayabu,Stacey,Neuhauser,Drake}, the cold front is steeper for higher impurity density and gentler for larger thermal diffusivity. The time required for the core $T_{\mathrm{e}}$ to decrease down to 120 eV governs the lifetime of the cold front because a linear instability occurs below 120 eV. \par
Such a behavior of the cold front is modeled by using the following thermal transport equation:
\begin{equation}
\dfrac{\partial T}{\partial t} = \chi \dfrac{\partial^{2} T}{\partial r^{2}} + \dfrac{\chi}{r}\dfrac{\partial T}{\partial r} - P(T - T_{l})(T - T_{h})^{2},
\label{nagumo_model}
\end{equation}
where $T_{l}$ and $T_{h}$ are the roots of the reaction term ($T_{h}>T_{l}$), and $P\left( >0\right)$ determines the magnitude of the net radiation losses. When the position of the cold front is far from the axis $r=0$, a traveling wave solution $\psi(r-ct)$ satisfies \cite{Witelski_2000}
\begin{equation}
\chi \dfrac{\partial^{2} \psi}{\partial x^{2}} + \left(c+ \dfrac{\chi}{R \left( t \right)} \right)\dfrac{\partial \psi}{\partial x} - P(\psi - T_{l})(\psi - T_{h})^{2}=0,
\label{large_radius_eq}
\end{equation}
where $x = r - ct$ denotes the reference frame moving with the velocity of the traveling wave $c$ and $R\left( t \right)$ denotes the radial position of the cold front. Equation (\ref{large_radius_eq}) has an exact solution for the wave shape in terms of a sigmoid function such that
\begin{equation}
\psi \left( x\right)= \dfrac{T_{h} - T_{l}}{1 + \exp[\lambda (x - x_{0})]} + T_{l},
\label{sigmoid_temperature}
\end{equation} 
with the traveling velocity $c$ \footnote{By determining $P$ such that the area of the reaction term between $T_{l}$ and $T_{h}$ matches that in Fig. 1, we obtain $-\sqrt{\dfrac{\chi P}{2}}(T_{h}-T_{l})\approx -0.21\ \mathrm{m/ms}$, where $T_{h}=120\ \mathrm{eV}$, $T_{l}=7.5\ \mathrm{eV}$, and $\chi=1\ \mathrm{m^{2}/s}$. The velocity of the cold front in Fig.~\ref{1e19_woheating} is on a similar timescale to this analytical result, although the assumptions are different.} and the parameter $\lambda$,
\begin{equation}
c=-\sqrt{\dfrac{\chi P}{2}}(T_{h}-T_{l})-\dfrac{\chi}{R},
\label{wave_velocity}
\end{equation} 
\begin{equation}
\lambda = \left( T_{h}-T_{l}\right) \sqrt{\dfrac{P}{2 \chi}},
\end{equation} 
where $x_{0}$ represents the inflection point. The parameter $\lambda$ is a key factor that determines the steepness of the temperature gradient. When Ohmic heating is negligible, the cubic term can be regarded as representing the curve of  $L \left(T_{\mathrm{e}} \right)$; therefore, $P \propto n_{\mathrm{z}}$ and $\lambda \propto \sqrt{n_{\mathrm{z}}/\chi}$, which indicates that the shape of the cold front is determined by $n_{\mathrm{z}}$ and $\chi$. The competition between the impurity radiation and thermal conduction, which  leads to the formation of a front with a steep gradient (large $\lambda$) when the radiation loss is dominant, is consistent with the analysis of Eq. (\ref{gradient_eq}) and the linear stability theory.  
%
%
%
%
%
%
%
%
%
\section{Electrothermal System}
\label{sec3}
\begin{figure}
  \centering
   \includegraphics[trim=0cm 0.1cm 0 0cm, clip,width=0.52\textwidth]{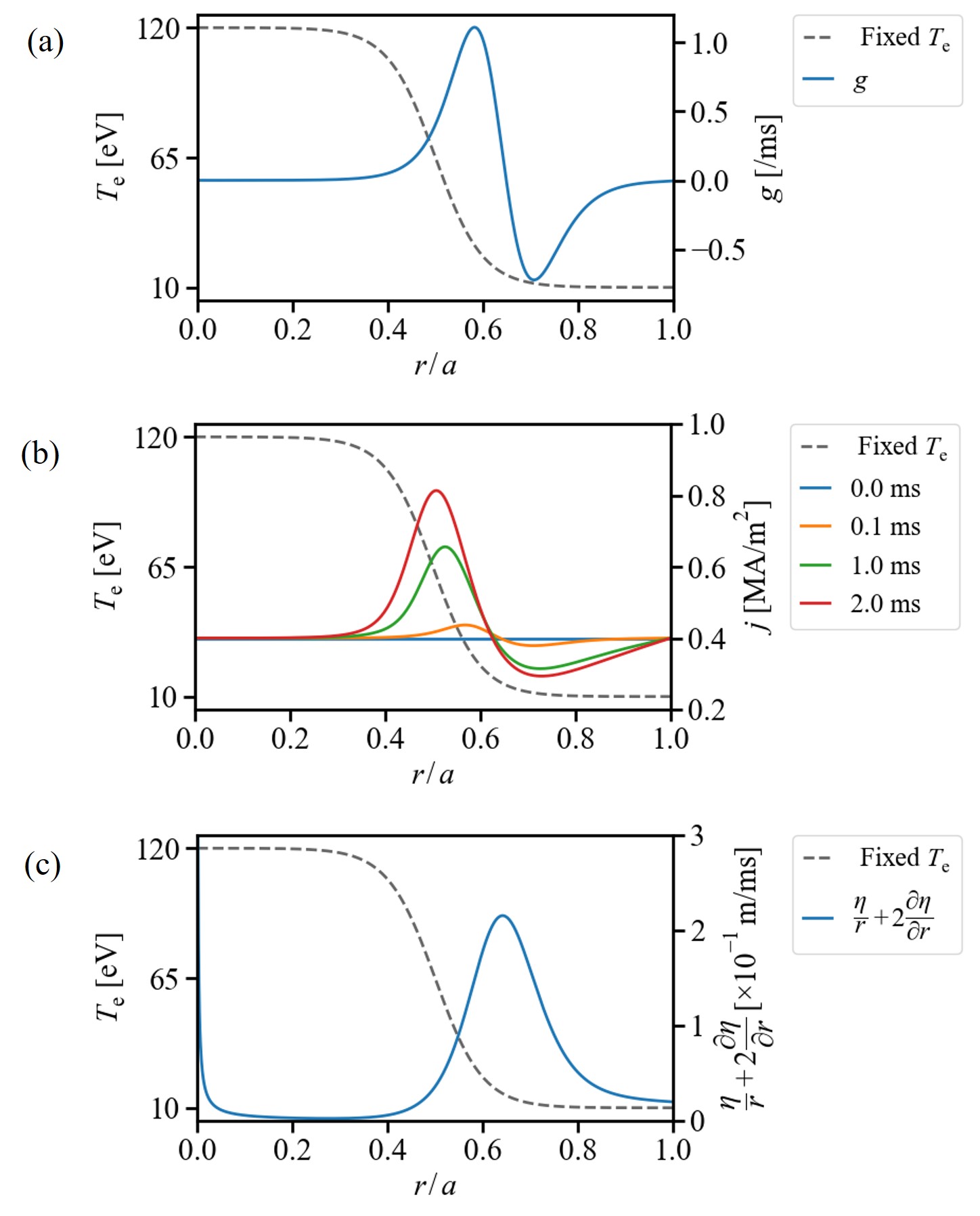}
  \caption{(a): Profile of the reaction term divided by the current density, $g \left(T_{\mathrm{e}}\right)$, (b): the time evolution of the current density profile obtained by numerically solving Eqs. (\ref{current_diffusion_eq_1}) and (\ref{current_diffusion_eq_2}), and (c): the profile of the advection coefficient for a fixed sigmoid temperature profile given by $T_{\mathrm{e}} = \dfrac{T_{h} - T_{l}}{1 + \exp[\lambda \left( r - r_0 \right)]} + T_{l}$, where $\lambda a = 20$, $r_{0}/a = 0.5$, $T_{l} = 10\ \mathrm{eV}$, $T_{h} = 120\ \mathrm{eV}$, and $Z_{\mathrm{eff}}=1.5$.}
  \label{current_diffusion_numerical_result_1}
\end{figure}
\begin{figure*}
  \centering
   \includegraphics[trim=0cm 0cm 0 0cm, clip,width=\textwidth]{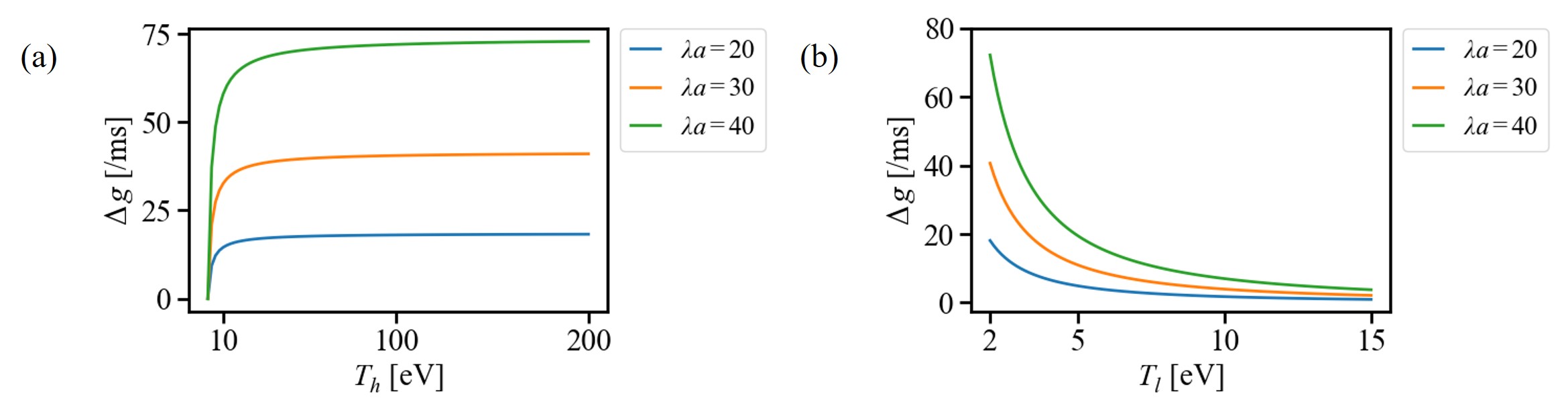}
  \caption{(a) Relationship between $\Delta\, g$ and the maximum electron temperature for each $\lambda$, where $T_{l}=2\ \mathrm{eV}$, and $a/r = 2$. (b) Relationship between $\Delta\, g$ and the minimum electron temperature for each $\lambda$, where $T_{h}=120\ \mathrm{eV}$, and $a/r = 2$. }
  \label{delta_g_Ts}
\end{figure*}
\begin{figure}
  \centering
   \includegraphics[trim=0cm 0cm 0 0cm, clip,width=0.52\textwidth]{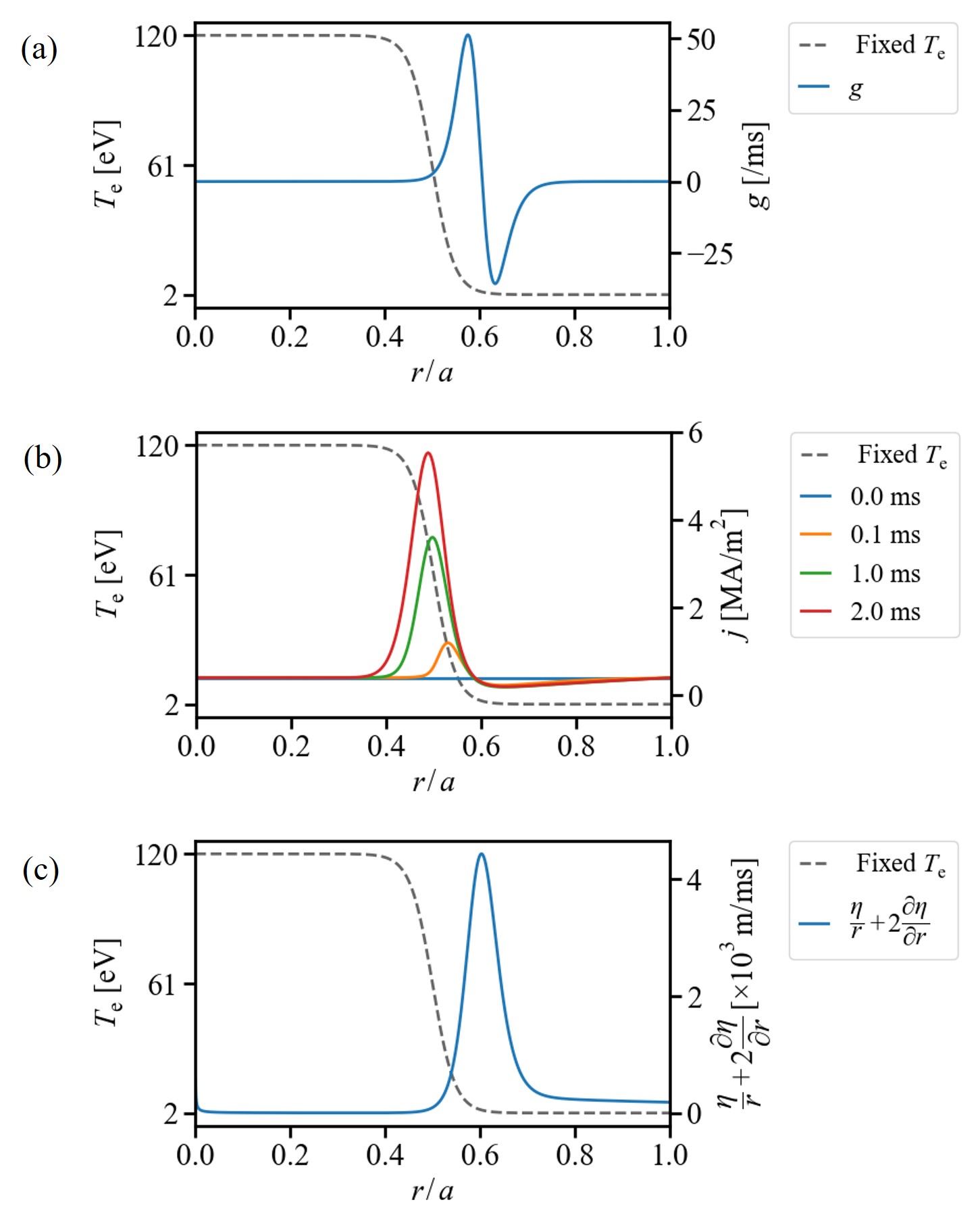}
  \caption{(a): Profile of the reaction term divided by the current density, $g \left(T_{\mathrm{e}} \right)$, (b): the time evolution of the current density profile obtained by numerically solving Eqs. (\ref{current_diffusion_eq_1}) and (\ref{current_diffusion_eq_2}), and (c): the profile of the advection coefficient for a fixed sigmoid temperature profile given by $T_{\mathrm{e}} = \dfrac{T_{h} - T_{l}}{1 + \exp[\lambda \left( r - r_{0} \right)]} + T_{l}$, where $\lambda a = 43$, $r_{0}/a = 0.5$, $T_{l} = 2\ \mathrm{eV}$, $T_{h} = 120\ \mathrm{eV}$, and $Z_{\mathrm{eff}}=1.5$.}
  \label{current_diffusion_numerical_result_2}
\end{figure}
The current diffusion is also described by a nonlinear reaction-diffusion (NRD) model. The time evolution of the current density in the cylindrical tokamak plasma can be expressed as
\begin{equation}
\mu \dfrac{\partial j}{\partial t}
 = \dfrac{1}{r}\dfrac{\partial}{\partial r}
   \!\left(  r\,\dfrac{\partial E}{\partial r} \right),
\end{equation}
where $\mu$ is the vacuum permeability and $E$ is the electric field parallel to the magnetic field. By using Ohm's law $E = \eta j$ and Spitzer resistivity $\eta \propto T_{\mathrm{e}}^{-3/2}$, the following current diffusion equation is derived.
\begin{equation}
\dfrac{\partial j}{\partial t}=\eta \dfrac{\partial^{2} j}{\partial r^{2}}+\left(\dfrac{\eta}{r} + 2\dfrac{\partial \eta}{\partial r}\right)\dfrac{\partial j}{\partial r}+g\left(T_{\mathrm{e}} \right)j,
\label{current_diffusion_eq_1}
\end{equation}
\begin{equation}
g\left( T_{\mathrm{e}} \right)=\dfrac{3}{2}\eta \left[\dfrac{5}{2}\left( \dfrac{\partial \ln T_{\mathrm{e}}}{\partial r} \right)^{2}-\dfrac{1}{r}\dfrac{\partial \ln T_{\mathrm{e}}}{\partial r}-\dfrac{1}{T_{\mathrm{e}}}\dfrac{\partial^{2} T_{\mathrm{e}}}{\partial r^{2}} \right],
\label{current_diffusion_eq_2}
\end{equation}
where $\mu$ is omitted for notational simplicity; note that it is retained in the following numerical results.
In deriving Eqs. (\ref{current_diffusion_eq_1}) and (\ref{current_diffusion_eq_2}), the gradient of the effective charge $Z_{\mathrm{eff}}$ is neglected. This NRD form of the current diffusion equation clarifies that, in addition to pure radial diffusion, the gradient of electrical resistivity causes current advection, and the reaction term $g\left(T_{\mathrm{e}} \right)j$ causes local drive and/or damp depending on the first and second radial derivatives of the electron temperature. Note that $g\left(T_{\mathrm{e}} \right)j$ depends linearly on the electric field $E=\eta j$; therefore, the reaction term can be significant as the current density is higher and the electron temperature is lower.\par
We illustrate the evolution of the current density profile when the electron temperature profile approaches the traveling wave solution, as described in Eq. (\ref{sigmoid_temperature}). Figure \ref{current_diffusion_numerical_result_1} (a) shows the profile of the reaction term divided by the current density, $g \left(T_{\mathrm{e}}\right)$, for a given $T_{\mathrm{e}}$ profile. Here, the minor radius $a$ is set to 1 m. It is seen from Fig. \ref{current_diffusion_numerical_result_1} (a) that the steep temperature gradient at $r/a = 0.6$ drives the current density, whereas the reaction term depletes the current density in the region where the electron temperature profile exhibits a pronounced concave-down curvature, as can be seen from Eq. (\ref{current_diffusion_eq_2}). Figure \ref{current_diffusion_numerical_result_1} (b) shows the time evolution of the $j$ profile obatined by numerically solving Eqs. (\ref{current_diffusion_eq_1}) and (\ref{current_diffusion_eq_2}) when the same $T_{\mathrm{e}}$ profile is fixed as that shown in Fig. \ref{current_diffusion_numerical_result_1} (a). A Dirichlet boundary condition is imposed at $r=a$, $E(a,t)=\eta(a)j(a,0)$, resulting in the non-conservation of the total current. The initial $j$ profile is radially uniform at $j=0.4\ \mathrm{MA/m^{2}}$. From 0 to 0.1 ms, the reaction term drives the current density inside $r/a = 0.6$ and decreases it outside $r/a = 0.6$. The perturbation of the current density grows due to the reaction term from 0.1 to 2.0 ms. Although $T_{\mathrm{e}}$ profile is here fixed, the peak of the current density gradually shifts toward the higher temperature region due to the advection term. Figure. \ref{current_diffusion_numerical_result_1} (c) shows the profile of the advection coefficient. The peak is located around $T_{l}$ in the cold front. Since $\dfrac{\partial \eta}{\partial r}=-\dfrac{3}{2}\dfrac{\eta}{T_{\mathrm{e}}}\dfrac{\partial T_{\mathrm{e}}}{\partial r}$, the advection coefficient becomes more significant at lower electron temperatures and larger temperature gradients. Therefore, the result shown in Fig. \ref{current_diffusion_numerical_result_1} (c) is reasonable. However, Fig. \ref{current_diffusion_numerical_result_1} (b) indicates that, due to diffusion from the peak of the current density, the negative current density gradient, $\partial j / \partial r < 0$, around $\rho = 0.65$ in the dip does not shift, even though the advection coefficient peaks at this location. Consequently, while the waveform is not preserved, the peak of the current density penetrates into the high-temperature region. In the actual evolution of the coupled system of $T_{\mathrm{e}}$ and $j$, the cold front itself moves at the traveling velocity of Eq. (\ref{wave_velocity}) and the current density perturbation protruding from the current profile propagates radially inward. In an earlier work \cite{Morozov_2005}, such a peaked current density perturbation is referred to as \textit{shark-fin like currents}, which we also call them hereafter. Shark-fin currents were also studied in connection with runaway current generation \cite{Putvinski_1997,Feher_2011, Pisztai_2022}. The formation and propagation of shark-fin currents was discussed as soliton-like solutions in \cite{Putvinski_1997}, in our viewpoint, which can be understood by considering them as an interplay of the propagation of the cold front [Eq. (\ref{sigmoid_temperature})] and the reaction term of the current diffusion equation [Eq. (\ref{current_diffusion_eq_2})]. \par
Because the local radiative collapse can nonlinearly destabilize ideal and/or resistive MHD instabilities through current density perturbations, the key plasma parameters governing their growth are of interest. We define a metric to estimate the current density gradient between shark-fin current and the dip, as shown in Fig. \ref{current_diffusion_numerical_result_1} (b). This metric is defined as $\Delta\, g  \coloneqq  g_{\mathrm{max}} - g_{\mathrm{min}}$, where $g_{\mathrm{max}}$ and $g_{\mathrm{min}}$ are the maximum and minimum values over the entire radial profile. The parameter $g$ can also be interpreted as the growth rate of the current density perturbation when the reaction term is dominant. When a flat initial current density profile is assumed, the behavior at the early stage of linear growth is described by
\begin{equation}
j=j_{0}\exp\left[g \left(T_{\mathrm{e}}\right)t\right],
\label{growth_rate}
\end{equation}
where $j_{0}$ is the initial current density. This expression indicates that $g$ represents the growth rate of the current density perturbation when the reaction term is dominant. For $g > 0$, the current density grows and forms a shark-fin current, whereas for $g < 0$ the current density decays and forms a dip. The difference of the current densities at $g_{\mathrm{max}}$ and $g_{\mathrm{min}}$, denoted $\Delta\, j$, is given by
\begin{equation}
\Delta\, j = j_{0} \left( \exp \left[g_{\mathrm{max}} t\right]-\exp \left[g_{\mathrm{min}} t\right] \right).
\label{amplitude}
\end{equation}
Figures \ref{delta_g_Ts} shows how $\Delta\, g$ varies with the temperature difference at the cold front interface, where $T_{h}$ and $T_{l}$ in Eq. (\ref{sigmoid_temperature}) are varied in Figs. \ref{delta_g_Ts} (a) and (b), respectively. Here, the radial curvature in the reaction term is assumed to be constant, with $a/r = 2$. Figure \ref{delta_g_Ts} (a) shows that, on the curve of constant $\lambda a$, $\Delta\, g$ increases with an increase in the maximum temperature. This result suggests that the magnitude of the current density perturbation depends on the impurity species because $T_{h}$ depends on the  radiative properties of the species. Figure \ref{delta_g_Ts} (b) shows that lower $T_{l}$ and larger $\lambda$ values yield higher magnitudes of the current density perturbation, which suggest that local radiative cooling amplifies the current density perturbation. To illustrate the above trends, we present another example of the time evolution of $j$ using $T_{l}=2$ eV and $\lambda a = 43$. Here, $\lambda a = 43$ is obtained by assuming that the average maximum temperature gradient of the cold front observed in the INDEX simulation is equivalent to the maximum temperature gradient at the inflection point of the sigmoid-type profile \footnote{The average maximum temperature gradient is calculated over 6.0-8.0 ms by taking the time average of the maximum gradient of the cold front shown in Fig. \ref{1e19_woheating} in the next section.}. Figure. \ref{current_diffusion_numerical_result_2} (a) shows that $\Delta\, g$ is more than ten times as large as that in Fig. \ref{current_diffusion_numerical_result_1} (a). Similarly, the peak of the advection coefficient is ten thousand times larger than that shown in Fig. \ref{current_diffusion_numerical_result_1} (c). Figure. \ref{current_diffusion_numerical_result_2} (b) shows that the shark-fin current at 2.0 ms is about six times larger than that in Fig. \ref{current_diffusion_numerical_result_1} (b). Here, the Dirichlet boundary condition, which is the same as that used in Fig. \ref{current_diffusion_numerical_result_1} (b), is applied at the boundary: $E(a,t)=\eta(a)j(a,0)$.\par 
Electrothermal dynamics around the electron temperature roots, where Ohmic heating and impurity radiation are balanced, can be considered from the above analyses. In the region where the growth rate $g$ is negative, a decrease in the current density results in a reduction of the Ohmic heating. This reduction in Ohmic heating locally leads to a decrease in electron temperature due to impurity radiation. The resulting temperature drop, in turn, increases the magnitude of the growth rate $g$ and causes the current density to decrease more rapidly. In the case of Fig. \ref{1e19_L-H_T}, the low-temperature stable root is located at 7.5 eV. After the radiative collapse, however, the stable solution on the low-temperature side of the cold front settles at a temperature signiﬁcantly lower than 7.5 eV, as a result of the depletion of the current density behind the cold front. In contrast, the shark-fin current formed by the positive growth rate $g$ leads to local reheating, which reduces the growth rate. When considering the net effect of the reaction term described by Eq. (\ref{current_diffusion_eq_2}), the magnitude of the current density must also be taken into account. Based on these electrothermal dynamics, the change in Ohmic heating cannot be neglected once a steep cold front has formed.\par

%
%
%
%
%
%
\section{Comparison with Tokamak Transport Simulation} 
\label{INDEX_results}
\subsection{The INDEX code}
We illustrate the electrothermal dynamics between cold front formation and current density perturbations via  one-dimensional simulations of the radiative collapse using the tokamak transport code INDEX \cite{INDEX}. The code solves 1D transport equations for electrons and multiple ion species in magnetic flux coordinates and a 2D MHD equilibrium using the Grad-Shafranov equation. The particle balance equations for each ion species are solved using the ionization and recombination rates extracted from the OpenADAS database  \cite{ADAS}. We use the magnetic surface label $\rho$ as a radial coordinate, which is defined as the toroidal magnetic flux normalized by its value at the plasma surface. In the region $\rho>1$, a simplified SOL model is applied \cite{Honda_2006}. The following simulations are performed with an up-down symmetric tokamak equilibrium with the major radius of 3 m, the plasma current is 1.3 MA, the toroidal magnetic field of 3 T, the minor radius of 1 m, the ellipticity of 1.6, and the triangularity of 0.3. For the boundary condition for describing electromagnetic coupling of a tokamak plasma with the conducting structure, the conducting wall of a conformal shape is placed at $r_{\mathrm{w}}/a= 1.3$ with a resistivity of $10^{-7}\ \Omega\, \mathrm{m}$, which defines the boundary condition for the current diffusion equation. Figure \ref{initial_profile} shows the initial profiles of the electron temperature and the current density. The initial profile of the ion temperature is the same as that of the electron temperature. The initial deuterium density at the plasma center is set to $5 \times 10^{19}\ \mathrm{m^{-3}}$. Neither auxiliary nor alpha heating are included. \par
\subsection{Model for thermal quench}
When setting our simulation of radiative collapse leading to disruptions, we need to consider its relationship to thermal quench. One scenario of natural disruption, typically in the carbon wall device, is that MHD instability precedes
the disruption, and the power and particle ﬂuxes to the divertor can increase peripheral impurities \cite{Yokoyama_APS_2023}. In the case of disruption mitigation, if the amount of injected impurities is sufficient and its penetration is deep enough, the entire core region can cool even without MHD. If the amount of impurities is small, both the enhanced heat transport due to MHD and the impurity radiation play roles in plasma cooling. Which occurs first varies depending on the injection scenarios, and multiple current spikes are often observed, in particular when the impurity radiation is insufficient \cite{Yokoyama_APS_2023,temperature_profile_shrinkage}. It is not the purpose of this paper to treat these scenarios systematically.\par
Alternatively, this paper focuses on the thermal instability of intermediate temperature plasmas around 120 eV with neon impurities. Although the setting is somewhat idealized to match our theoretical framework, it provides a useful case for physical understanding. The radiative collapse is modeled using an artificial impurity source and enhanced thermal conduction. While no initial background impurity is included when the simulation starts from $t=0$ ms, the neon density is increased from 0.5 to 0.8 ms using an artificial source term to achieve a spatially uniform impurity profile at a neon density of $10^{19}\ \mathrm{m^{-3}}$. The anomalous electron thermal transport, which mimics the breaking of the magnetic surfaces \cite{thermal_quench}, is also modeled by imposing a high thermal diffusivity uniformly. A thermal diffusivity of $200\ \mathrm{m^{2}/s}$ is used for the thermal quench duration that is set for each simulation from $t=0.55$ ms, and of $1\ \mathrm{m^{2}/s}$ otherwise. These settings allow us to observe how radiative collapse occurs in a plasma with $T_{h} \simeq 120$ eV, in which a given amount of
impurities is assimilated.\par
\begin{figure}
  \centering
  \includegraphics[trim=0cm 0.1cm 0 0cm, clip,width=0.5\textwidth]{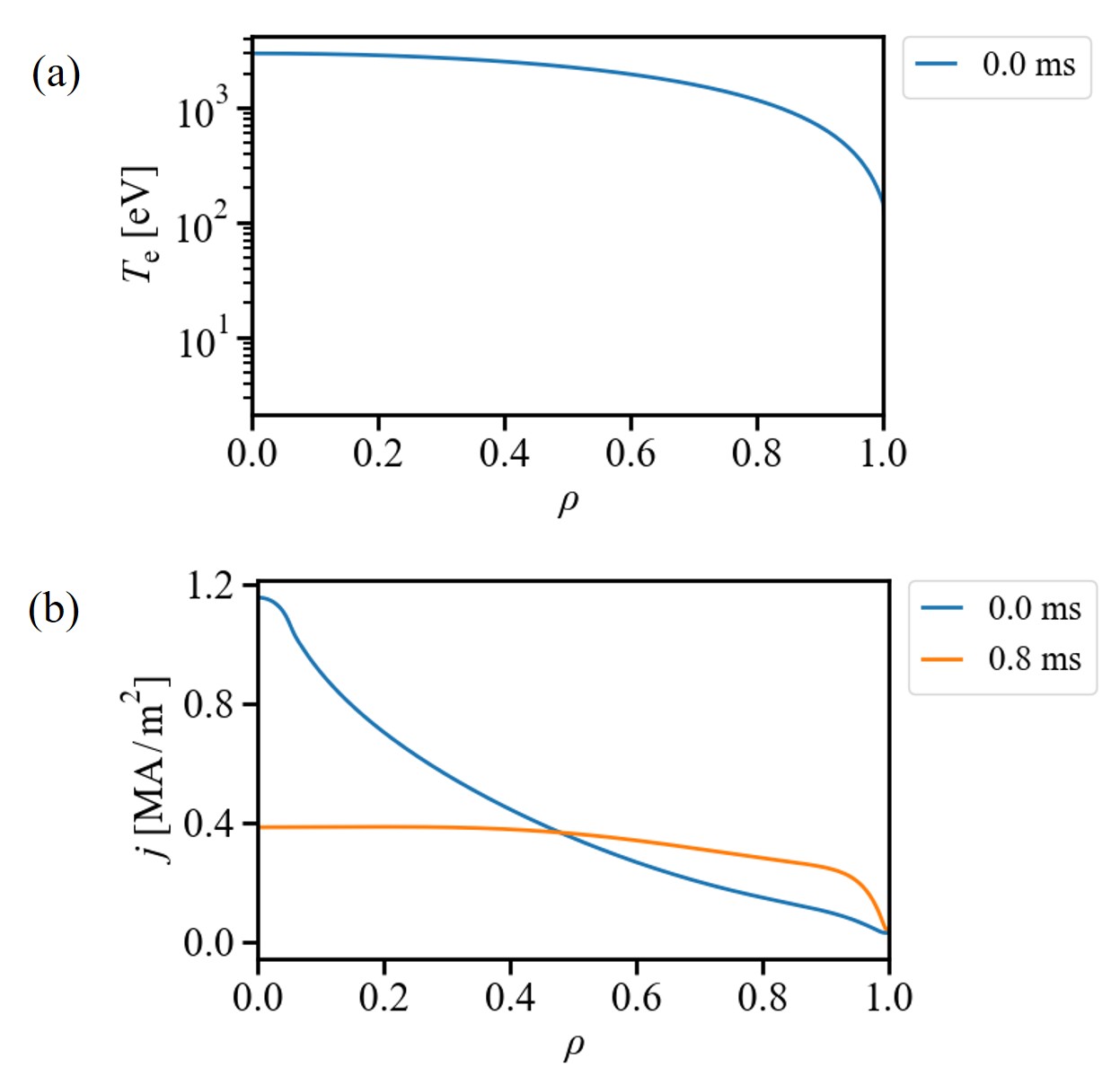}
  \caption{(a) Initial profile of the electron temperature. (b) Current density profiles: initial one and one at the time when hyper-resistivity is turned off in the simulation case with the neon density of $10^{19}\ \mathrm{m^{-3}}$.}
  \label{initial_profile}

  \includegraphics[trim=0cm 0cm 0 0cm, clip,width=0.5\textwidth]{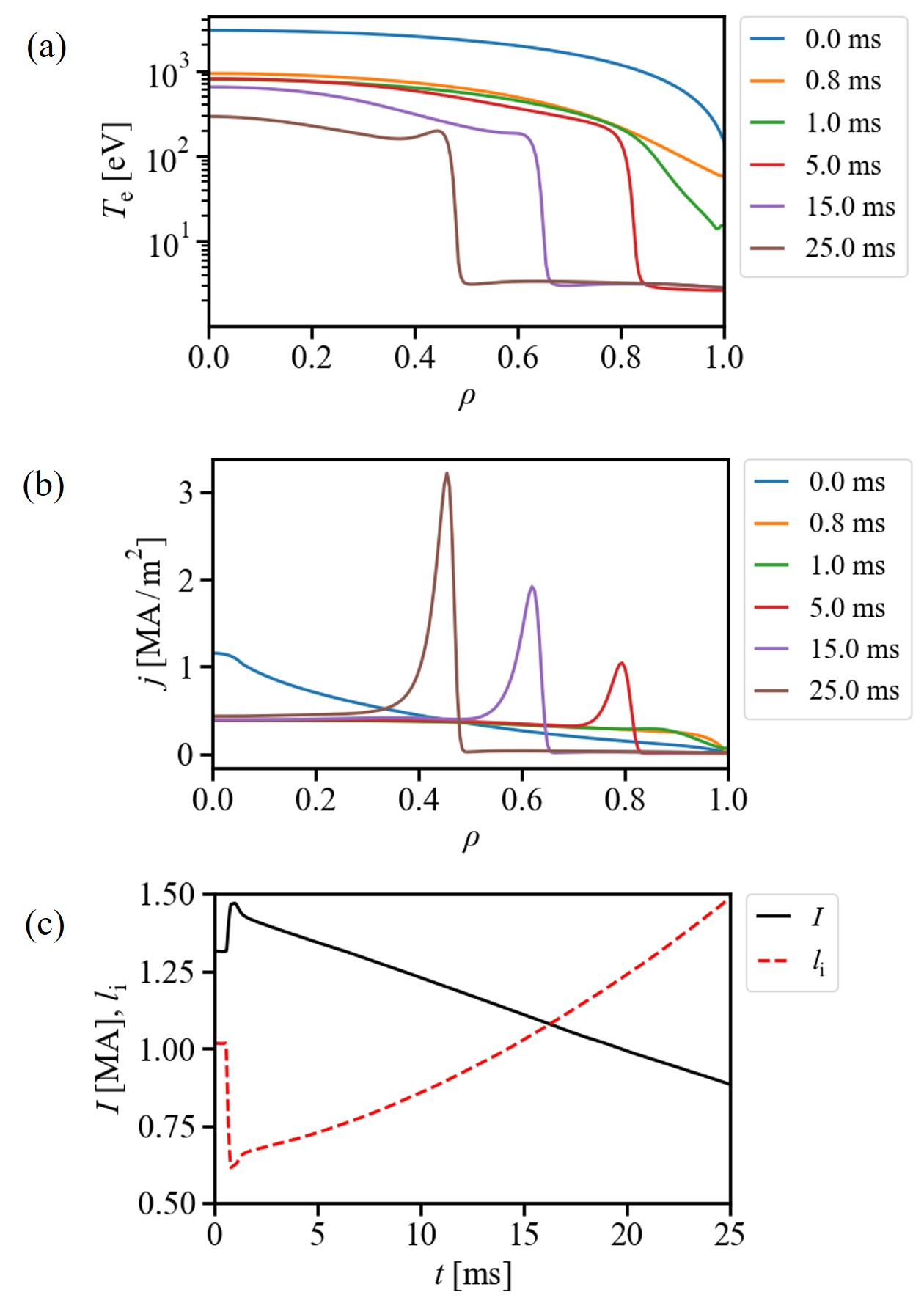}
  \caption{Time evolution of (a) the electron temperature profile, (b) the current density profile, and (c) the plasma current and the internal inductance for $n_{\mathrm{z}} = 10^{19}\ \mathrm{m^{-3}}$.}
  \label{1e19_modeling}
\end{figure}

The hyper-resistivity \cite{Ward_1989} is applied to flatten the current profile. The following tangent hyperbolic profile, similar to that in \cite{Narden_2023}, is used:
\begin{equation}
\lambda_{h}=\dfrac{C_{0}}{2}\left(1+\tanh \dfrac{C_{1}-\rho}{C_{2}}\right)
\label{hyper-resistivity}.
\end{equation}
The hyper-resistivity is turned on at 0.55 ms and turned off at 0.8 ms, with parameters $C_{0}=4.69,\ C_{1}=0.9$, and $C_{2}=0.05$. Figure \ref{initial_profile} (b) shows the current density profile at 0.8 ms when the hyper-resistivity is turned off.\par
\begin{figure}
  \includegraphics[trim=0cm 0cm 0 0cm, clip,width=0.5\textwidth]{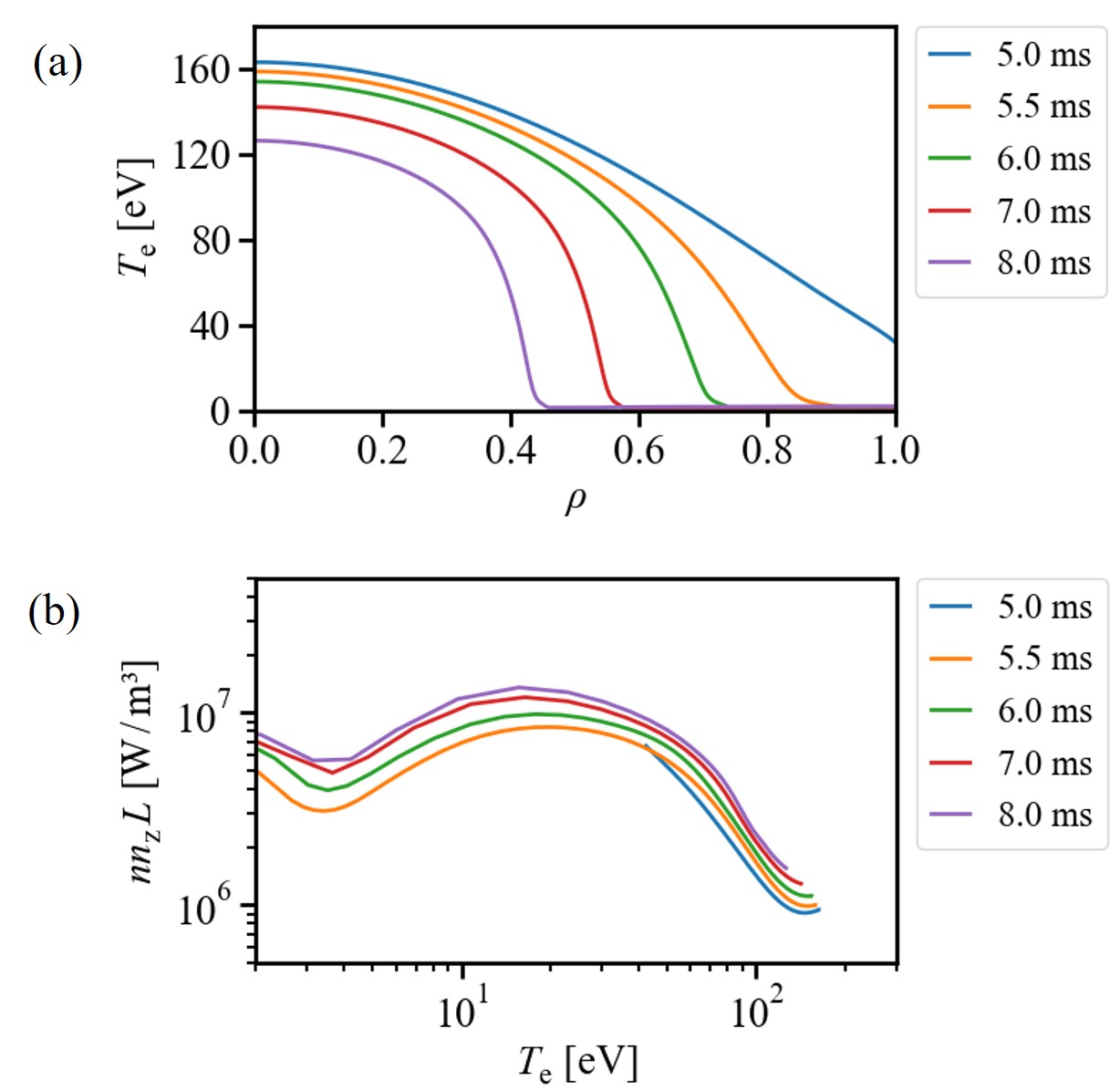}
  \caption{Time evolution of (a) the electron temperature profile and (b) the corresponding relationship between the neon radiation and electron temperature for $n_{\mathrm{z}} = 10^{19}\ \mathrm{m^{-3}}$ in the case where the Ohmic heating is turned off. The thermal diffusivity is $200\ \mathrm{m^{2}/s}$ from 0.55 to 5 ms. In all other time regions, the thermal diffusivity is $1\ \mathrm{m^{2}/s}$.}
  \label{1e19_woheating}
\end{figure}
  \begin{figure}
  \includegraphics[trim=0cm 0cm 0 0cm, clip,width=0.5\textwidth]{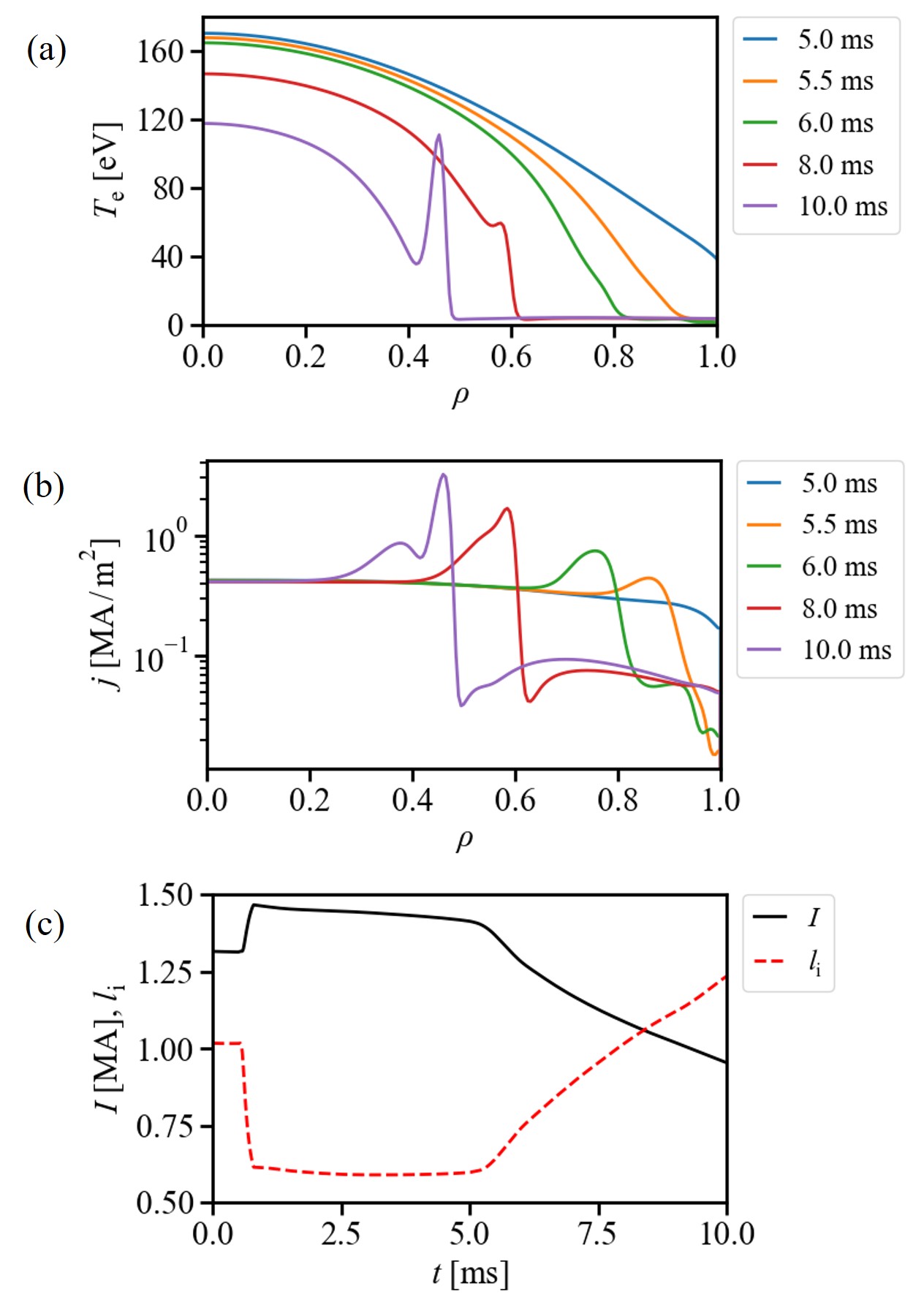}
  \caption{Time evolution of (a) the electron temperature profile, (b) the current density profile, and (c) the plasma current and the internal inductance for $n_{\mathrm{z}} = 10^{19}\ \mathrm{m^{-3}}$. The thermal diffusivity is $200\ \mathrm{m^{2}/s}$ from 0.55 to 5 ms. In all other time regions, the thermal diffusivity is $1\ \mathrm{m^{2}/s}$.}
  \label{1e19_profiles}
\end{figure}
\begin{figure*}[]
  \includegraphics[width=0.7\textwidth]{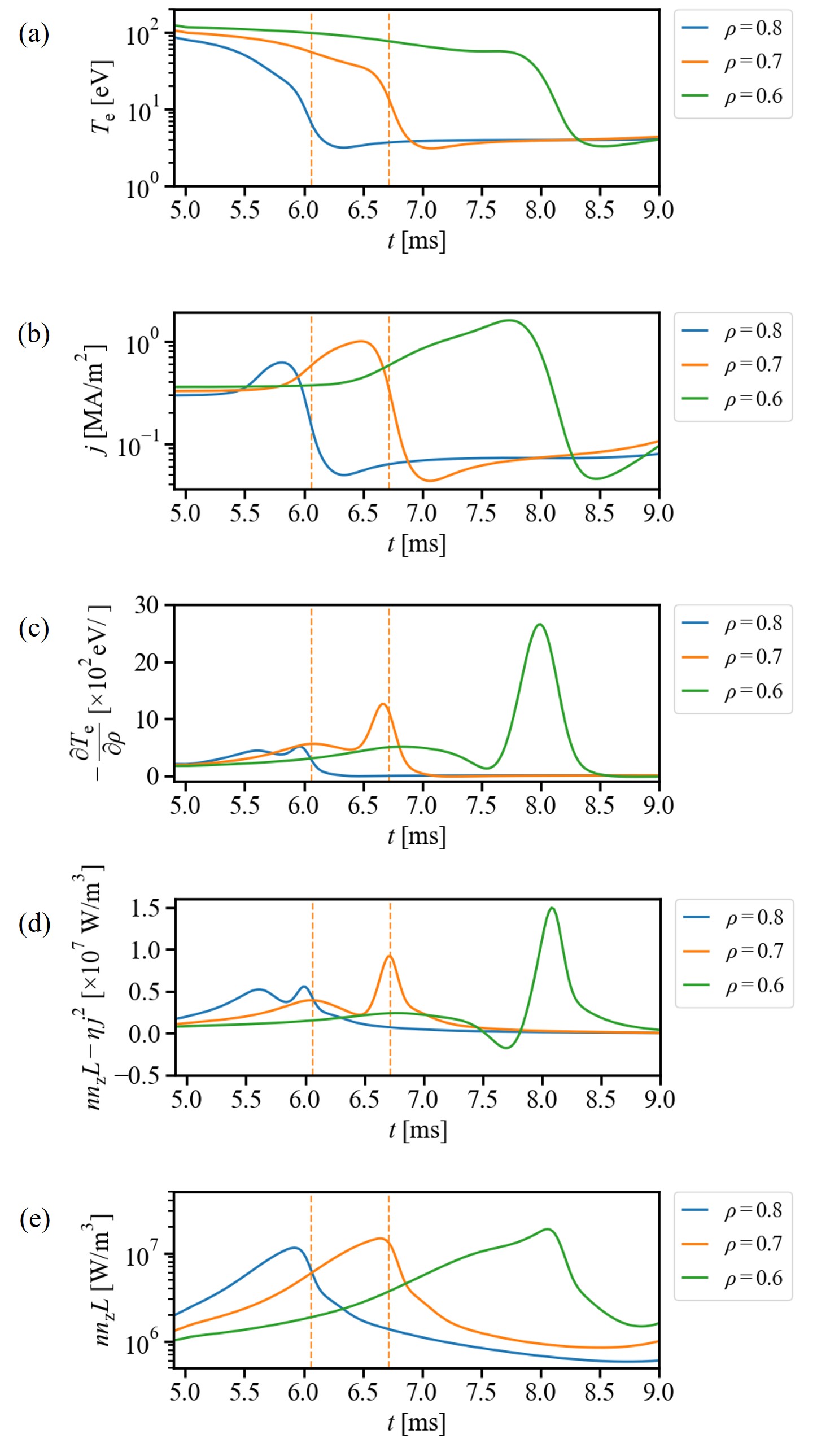}
  \caption{Time evolution of (a) the electron temperature, (b) the current density, (c) the electron temperature gradient, and (d) the net radiation, which is calculated by subtracting the Ohmic heating from the neon radiation. (e) Neon radiation at $\rho = 0.8, 0.7$ and 0.6, for $n_{\mathrm{z}} = 10^{19}\ \mathrm{m^{-3}}$. The dashed lines in the left panels indicate the times at which the net radiation peaks at $\rho = 0.7$.}
  \label{1e19_time_evolution}
\end{figure*}
\begin{figure}
  \includegraphics[trim=0cm 0cm 0cm 0cm, clip, width=0.52\textwidth]{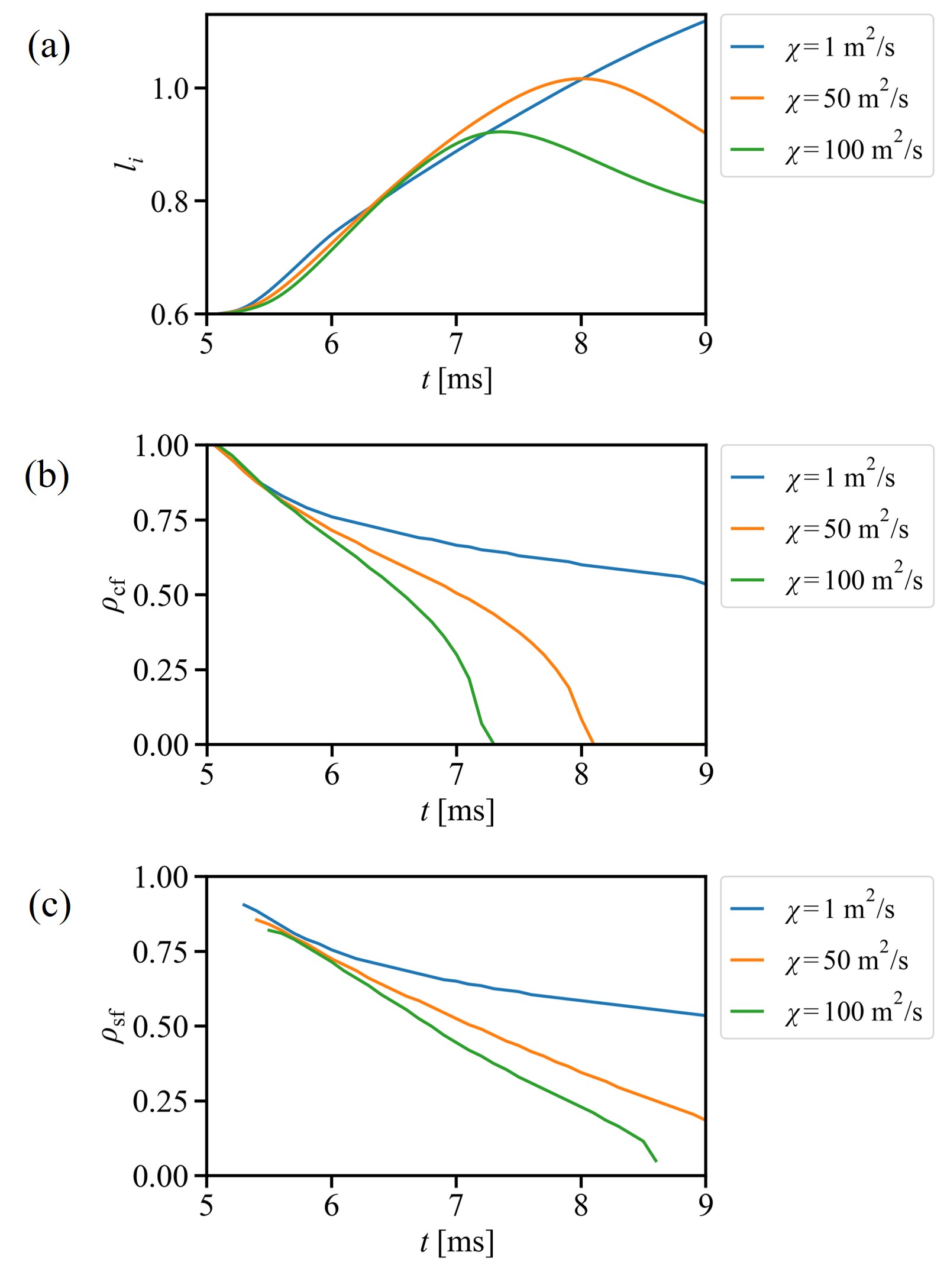}
  \caption{Time evolution of (a) the internal inductance, (b) the position of the cold front $\rho_{\mathrm{cf}}$, and (c) the position of the shark-fin current $\rho_{\mathrm{sf}}$ for $\chi = 1$, 50, and 100 $\mathrm{m^{2}/s}$.}
  \label{internal_induc_and_kai}
\end{figure}
%
%
%
%
%
%
%
\subsection{Simulation results}
\label{simulation_results}
Figure \ref{1e19_modeling} shows the time evolution of the electron temperature and current density profiles, the plasma current, and the internal inductance in the case where Ohmic heating is included. Here, the vertical axis $j$ in Fig. \ref{1e19_modeling} (b) represents the component of the current density parallel to the magnetic field, defined as $ j = \frac{\left\langle \mathbf{J} \cdot \mathbf{B} \right\rangle}{\langle B^{2} \rangle} B_{0}$, where $\langle \cdot \rangle$ denotes the flux-surface average and $B_{0}$ is the magnetic field strength at the magnetic axis. The same definition is used in the subsequent figures. In this case, the thermal quench is set to turn on at 0.55 ms and turn off at 0.8 ms. In Figure \ref{1e19_modeling} (a), the electron temperature profile outside $\rho=0.8$ collapses after 0.8 ms because the heat flow
from the inner core stops. A cold front is then formed at 5.0 ms and propagates toward the plasma center between 5.0 and 25.0 ms. In the simulation of Fig. \ref{1e19_modeling},  the neon impurity is presented radially uniform, and the cold front are not driven by the impurity transport; it is considered to be formed by the electron temperature dependence of the radiative cooling rate. The developed cold front is seen to accompany the shark-fin current [Fig. \ref{1e19_modeling} (b)], while the current density behind the shark-fin current is flattened near zero. In contrast, the shark-fin current grows significantly as it propagates inwardly,
which leads to an increase in the internal inductance, as shown in Fig. \ref{1e19_modeling} (c). In the simulation, the plasma current decays at the rate of 0.5 MA per 20 ms; such a relatively slow current decay accompanying the $l_{\mathrm{i}}$ rise is often observed experimentally, such as in the neon gas puffing discharges in JT-60U \cite{temperature_profile_shrinkage}. \par
The current perturbations lead to the electrothermal dynamics through Ohmic heating. The coupling between $T_{\mathrm{e}}$ and $j$ becomes more pronounced when the electron temperature after the thermal quench is low. To illustrate
such a situation, we use the same simulation condition as in the simulation of Fig. \ref{1e19_modeling}, whereas the thermal quench duration is extended to 5 ms to obtain a flat electron temperature profile around 120 eV. For comparison, Fig. \ref{1e19_woheating} (a) shows the time evolution of the electron temperature profiles in the case where the Ohmic heating is turned off. The neon radiation governs the temperature evolution after the thermal quench ($>$ 5 ms), and the temperature gradient becomes steeper below approximately 120 eV. The radiation power over the radial profile is shown as a function of $T_{\mathrm{e}}$ in Fig. \ref{1e19_woheating} (b). Because the temperature gradient is monotonic at each time in Fig. \ref{1e19_woheating} (a), each radiation peak around 20 eV in Fig. \ref{1e19_woheating} (b) corresponds to a location within the cold front. The radiation power sharply rises by an order of magnitude from the minimum value of approximately 120 eV. The trend of the curve below 120 eV  ($\partial nn_{z}L/\partial T_{\mathrm{e}} < 0$) is consistent with that under the coronal equilibrium shown in Fig. \ref{1e19_L-H_T}. \par
Figure \ref{1e19_profiles} (a) shows the time evolution of the electron temperature profile in the case where the Ohmic heating is turned on. While the cold front propagates toward the plasma center in a similar manner to that shown in Fig. \ref{1e19_woheating}, a spike in the $T_{\mathrm{e}}$ profile is formed at the inner radius of the cold front. Figure \ref{1e19_profiles} (b) shows the corresponding current density profiles. The magnitude of the current density peak increases from 5.5 to 10.0 ms. The logarithmic plot of Fig. \ref{1e19_profiles} (b) also shows a dip in the current density right behind the cold front. Figure \ref{1e19_profiles}(c) shows the time evolution of the plasma current and the internal inductance. While the plasma current and the internal inductance remain nearly constant after turning off the hyper-resistivity at 0.8 ms, the current decay and the $l_{\mathrm{i}}$ rise are driven by the radiative collapse once the thermal quench heat flow stops.\par
For a detailed comparison with our analytical model in Sec. \ref{sec3}, Fig. \ref{1e19_time_evolution} shows the time evolution of (a) the electron temperature, (b) the current density, (c) the electron temperature gradient, (d) the net radiation which calculated by subtracting the Ohmic heating from the neon radiation, and (e) the neon radiation at three radial position $\rho = 0.8, 0.7$ and 0.6. It can be seen that as the cold front propagates inward from $\rho = 0.8$ to 0.6 [Fig. \ref{1e19_time_evolution} (a)], the shark-fin current follows the cold front position and increases its magnitude [Fig. \ref{1e19_time_evolution} (b)]. The time evolution of the net radiation loss $nn_{\mathrm{z}}L-\eta j^{2}$ [Fig. \ref{1e19_time_evolution} (d)]  shows two peaks indicated by the vertical dashed lines at 6.1 and 6.7 ms on the curve of $\rho=0.7$, while the radiation power [Fig. \ref{1e19_time_evolution} (e)] shows only one peak at 6.7 ms. Interestingly, the behavior of the net radiation loss explicitly reflects the electron temperature gradient [Fig. \ref{1e19_time_evolution} (c)], which suggests that the electron temperature gradient is determined by the radial gradient of the reaction term, which includes both Ohmic heating and impurity radiation. The decrease in the net power loss after the first peak corresponds to an increase in the Ohmic heating due to the formation of a shark-fin current. The second peak at 6.7 ms occurs when the shark-fin current passes and the current density decreases at $\rho=0.7$. After 6.7 ms, therefore, the electron temperature curvature becomes concave-down, and the reaction term depletes the current density locally at $\rho = 0.7$; $j$ then reaches a minimum value of 0.04 $\mathrm{MA/m^{2}}$ at 7.0 ms. This current density perturbation leads to a decrease in the electron temperature to around 3 eV, although the stable temperature is approximately 7.5 eV at the current density immediately after the thermal quench, as shown in Fig. \ref{1e19_L-H_T}. The same overall trends are also observed at $\rho=0.6$ and $\rho=0.8$ following the inward propagation of the cold front.\par
The inward propagation of growing current perturbation causes a steep rise in the internal inductance. The dependence of the $l_{\mathrm{i}}$ rise on
the thermal diffusivity is thus of interest. Figure \ref{internal_induc_and_kai} shows the temporal evolution of (a) the internal inductance, (b) the position of the cold front $\rho_{\mathrm{cf}}$, and  (c) the position of the shark-fin current for $\chi = 1$, 50, and $100\ \mathrm{m^{2}/s}$ after the thermal quench. Here, $\rho_{\mathrm{cf}}$ is defined as the position of 30 eV, and $\rho_{\mathrm{sf}}$ is the position of the peak of the shark-fin current. Except for the thermal diffusivity after the thermal quench, all the simulation parameters are identical to those used in Fig. \ref{1e19_profiles}. From 5 to 6.5 ms, the internal inductance is larger for lower thermal diffusivity [Fig.\ref{internal_induc_and_kai} (a)], although the position of the shark-fin current is almost the same in all three cases [Fig. \ref{internal_induc_and_kai} (c)]. This trend arises because lower thermal diffusivity leads to the formation of a narrower cold front and a larger shark-fin current. Here, the temporally averaged width of the cold front, calculated as the distance between 30 eV and 120 eV over 5.5-6 ms, is $\Delta\, \rho = 0.27$, $0.32$, and $0.40$ for $\chi=1$, $50$, and $100\ \mathrm{m^{2}/s}$, respectively. Additionally, the temporally averaged peak of the shark-fin current over 5.5-6 ms is $0.60$, $0.47$, and $0.42\ \mathrm{MA/m^{2}}$ for $\chi=1$, $50$, and $100\ \mathrm{m^{2}/s}$, respectively. In contrast, from 6.5 to 7.5 ms, the internal inductance for $\chi = 50$ and $100\ \mathrm{m^{2}/s}$ becomes larger than that for $\chi = 1\ \mathrm{m^{2}/s}$ [Fig. \ref{internal_induc_and_kai} (a)]. This behavior is attributed to the faster propagation of the cold front and the shark-fin current at higher thermal diffusivity [Figs. \ref{internal_induc_and_kai} (b) and (c)]. As expected, the faster-propagating cold front reaches $\rho = 0$, and the high-temperature region in the core collapses more rapidly. Consequently, the internal inductance for $\chi = 50$ and $100\ \mathrm{m^{2}/s}$ starts to decrease at approximately 7.3 and 8 ms, respectively. The internal inductance for $\chi = 1\ \mathrm{m^{2}/s}$ at 9 ms exceeds the peak values for $\chi = 50$ and $100\ \mathrm{m^{2}/s}$ [Fig. \ref{internal_induc_and_kai} (a)], because the shark-fin current increases, as discussed in Sec. \ref{sec3}.\par 
%
%
%
%
%
%
%
%
\vline
\section{Discussion and Conclusion}
\label{dis_con}
In this work, the key parameters governing the current density perturbations induced by cold front formation and the resulting electrothermal dynamics have been investigated. The current diffusion equation, formulated as a reaction-diffusion model, explains that the current density perturbation can be formed by the first and second radial derivatives of the electron temperature profile when the plasma becomes highly resistive. The reaction term in the current diﬀusion equation also drives the growth of current density perturbation. For a monostable electron temperature profile whose edge region has collapsed due to impurity radiation, such as a sigmoid profile, a shark-fin current forms around the steep temperature gradient, whereas the current density behind the cold front decreases in the concave-down region of the electron temperature profile. While the shark-fin current can induce reheating through Ohmic heating, a decrease in current density can shift the stable root of the power balance towards the lower the temperature and further amplify the current density perturbation. The INDEX simulations indicate that the electron temperature and current density exhibit behavior consistent with the models presented in this work. \par
The edge radiation collapse due to impurities and the resultant current disturbance have been investigated in numerous simulation studies, including those employing sophisticated MHD models. Because of the limitations inherent in of the one-dimensional transport model, this work does not address the disruption processes in a self-consistent manner. Nevertheless, we emphasize that more advanced MHD models also include the reaction-diffusion mechanisms in the temperature and current density, whose role in the electrothermal dynamics associated with the dissipation of thermal and magnetic energy during tokamak disruptions deserves particular attention. We also note that in the literature \cite{Morozov_2005}, the pressure gradient is considered a possible mechanism for driving a shark-fin type current. While the present work focuses only on the currents driven by the resistivity profile effect, the impact of radiation collapse on the MHD equilibrium, including such pressure-driven currents, is of interest for further analysis.\par
According to our analysis, the propagation of a cold front from the plasma edge to the core and the amplification of a shark-fin current cause an increase of internal inductance. An increase $l_{\mathrm{i}}$ at the beginning of current decay, including the phase during thermal quench, has been observed in different classes of disruptions in JT-60U experiments \cite{Yokoyama_APS_2023,temperature_profile_shrinkage}. Because these experiments include high-$\beta$ discharges on the carbon-wall environment as well as intentional disruptions caused by neon gas puffing \cite{temperature_profile_shrinkage}, it is expected that the transport of impurities from the plasma edge contributes to the development of the cold front. Although the present analysis did not consider the $Z_{\mathrm{eff}}$ profile, the reaction-diffusion model developed here
is also applicable to analyze such experiments. \par
The insights gained from this study are also useful for understanding the thermal stability of tokamak plasma during disruption mitigation. One of the key challenges in developing disruption mitigation techniques is to increase the free electron density of the plasma while avoiding a significant increase in the population of target electrons for avalanching runaway electrons with
high-$Z$ atoms \cite{Hesslow_2019}. Therefore, it is necessary to obtain a better understanding of plasma conditions that are marginally stable against thermal collapse. The SPI simulation by INDEX \cite{INDEX} has shown that the impurity density front precedes the propagation of the cold front, which was recently confirmed by the experimental observations \cite{Bodner_2025}.  The physical understanding developed in this work is also useful for analyzing the thermal stability of the lukewarm plasma formed after the penetration of pure deuterium or deuterium-neon mixed pellets \cite{Narden_2020}, which will be discussed in future publications.
\section*{ACKNOWLEDGMENTS}
This work was supported in part by Grants-in-Aid for Scientific Research (MEXT KAKENHI Grant No. 25K00982) and JST SPRING, Grant Number JPMJSP2110.
\section*{Data Availability Statement}
The data that support the findings of this study are available from the corresponding author upon reasonable request.

\end{document}